\documentclass{article}
\pdfoutput=1 

\usepackage{arxiv}
\usepackage{abbreviations}

\usepackage[utf8]{inputenc} 
\usepackage[T1]{fontenc}    
\usepackage{hyperref}       
\usepackage{url}            
\usepackage{booktabs}       
\usepackage{amsfonts}       
\usepackage{nicefrac}       
\usepackage{microtype}      
\usepackage{lipsum}
\usepackage{graphicx}
\usepackage[square,sort,comma,numbers]{natbib}
\usepackage{amsmath}
\usepackage{amssymb}

\hypersetup{%
  colorlinks=true,
  linkcolor=blue,
  citecolor=blue,
  urlcolor=blue
}

\title{Studies of Evolved Stars in the next decade}

\author{
  Peter Scicluna\thanks{\texttt{peterscicluna@asiaa.sinica.edu.tw}}$^{\:\:^{1}}$~~$\bullet$~~
  Hiroko Shinnaga$^{2}$~~$\bullet$~~
  Jonathan Marshall$^{1}$~~$\bullet$~~
  Jan Wouterloot$^{3}$~~$\bullet$~~
  Iain McDonald$^{4}$\\
  \textbf{Steven Goldman$^{5}$~~$\bullet$~~
  Sofia Wallstr\"om$^{6}$~~$\bullet$~~
  Jacco Th. van Loon$^{7}$~~$\bullet$~~
  Thavisha Dharmawardena$^{1}$}\\
  \textbf{Lapo Fanciullo$^{1}$~~$\bullet$~~
  Sundar Srinivasan$^{8}$}\\\\
  \textit{$^{1}$Academia Sinica Institute of Astronomy and Astrophysics, AS/NTU Astronomy-Mathematics Building,}\\\textit{No 1. Sec. 4 Roosevelt Rd, Taipei, Taiwan}\\
  \textit{$^{2}$Department of Physics and Astronomy, Amanogawa Galaxy Astronomy Research Center (AGARC),}\\\textit{Kagoshima University, 1-21-35 Korimoto, Kagoshima, Japan}\\
  \textit{$^{3}$East Asian Observatory (JCMT), 660 N. A`ohoku Place, Hilo, Hawai`i, USA, 96720}\\
  \textit{$^{4}$Jodrell Bank Centre for Astrophysics,
  School of Physics and Astronomy,
  University of Manchester,}\\\textit{M13 9PL,
  Manchester, UK}\\
  \textit{$^{5}$Space Telescope Science Institute,
  3700 San Martin Drive, Baltimore,
  MD 21218, USA}\\
  \textit{$^{6}$Institute of Astronomy, KU Leuven,
  Celestijnenlaan 200D bus 2401,
  3001 Leuven, Belgium}\\
  \textit{$^{7}$Lennard-Jones Laboratories,
   Keele University,
   ST5 5BG, UK}\\
   \textit{$^{8}$Instituto de Radioastronom\'ia y Astrof\'isica, UNAM., Apdo. Postal 72-3 (Xangari),}\\\textit{Morelia, Michoac\'an 58089, Michoac\'{a}n, M\'{e}xico}\\
}

\begin{document}
\maketitle

\begin{abstract}
This white paper discusses recent progress in the field of evolved stars, primarily highlighting the contributions of the James Clerk Maxwell Telescope. 
It discusses the ongoing project, the \emph{Nearby Evolved Stars Survey} (NESS), and the potential to build upon NESS in the next decade.
It then outlines a number of science cases which may become feasible with the proposed 850\,$\mu$m camera which is due to become available at the JCMT in late 2022.
These include mapping the extended envelopes of evolved stars, including in polarisation, and time-domain monitoring of their variations.
The improved sensitivity of the proposed instrument will facilitate statistical studies that put the morphology, polarisation properties and sub-mm variability in perspective with a relatively modest commitment of time that would be impossible with current instrumentation.
We also consider the role that could be played by other continuum wavelengths, heterodyne instruments or other facilities in contributing towards these objectives.








\end{abstract}


\section{Introduction}
\label{sec:intro}








Evolved stars play key roles in the chemical enrichment of the Universe; most directly, they return material to the interstellar medium (ISM), fuelling future star formation with hydrogen, and enrich the ISM with dust and the products of nucleosynthesis \citep{Karakas2014IAUS..298..142K}. 
The rates at which this mass is lost are also critical; in low- and intermediate-mass stars, taking place primarily on the asymptotic giant branch (AGB), it impacts the amount of metals they process (He, C, N, O),  and s-process enrichment \citep{Karakas2014PASA...31...30K}.
Meanwhile for massive stars, mass-loss controls their lifetimes (and hence the injection of ionising photons into the ISM) and which supernova channel will consume them, dictating the production of iron-peak elements and the turbulence of the ISM \citep[e.g. see][for a recent review]{Meynet2017IAUS..329....3M}.
To understand the physics and chemistry of the ISM, it is therefore crucial to understand the physics of mass loss and its impact on stellar evolution and the ISM.

Nevertheless, a number of important questions remain open. 
We do not yet understand how, when and how much mass-loss takes place on the AGB, meaning that yields for low- and intermediate-mass stars are difficult to predict, as they are sensitive to the total length of the AGB phase. 
While the winds are expected to be driven by radiation pressure on dust, it is not clear when this becomes relevant 
and how it depends on the properties of the dust, whose formation process remains similarly poorly understood \citep[e.g.][]{Hofner2018A&ARv..26....1H}. 
For low-mass stars, mass loss on the red giant branch (RGB) has a significant impact on their AGB evolution and on the initial--final mass relation for white dwarves.
We know that mass-loss from red supergiants (RSGs) can dramatically alter the evolution of massive stars, controlling their lifetimes and dramatically altering their pre-supernova nucleosynthesis, yet the mechanisms driving the mass loss remain elusive with several competing suggestions \citep{Meynet2017IAUS..329....3M}.
In all of these cases, the time-variability of the mass loss will also play an important role, as will any companions, if present, which may shape the outflow \citep[e.g.][]{Kim2015,Kim2017} or entirely alter the evolutionary pathway \citep[e.g.][]{Smith2015}. 

Sub-mm astronomy has a key role to play in exploring these questions.
The low-J rotational transitions of CO are the most direct means of estimating gas mass-loss rates for evolved stars, while the $^{13}$CO isotopologues are an effective way of measuring the outcomes of nucleosynthesis when the photosphere is obscured by circumstellar dust and provide an independent constraint on the optical depth of the envelope \citep[e.g.][]{GreavesHolland1997}.
If the winds are dust-driven, the properties of the dust are key to driving the wind, and the sub-mm continuum is sensitive to both the size and composition of dust grains. 

\section{Current status of evolved-star science in the sub-mm}
\label{sec:current}







Single-dish sub-mm studies of evolved stars have a long and fruitful history.
As luminous sources with abundant molecules, CO-line emission from evolved stars has been a particular target.
This has allowed us to determine mass-loss rates for samples of objects \citep[e.g.][]{Young1995ApJ...445..872Y,Knapp1985ApJ...292..640K,Kahane1994A&A...290..183K}.
While many observations of the low-J CO lines exist, homogeneous compilations for large, volume-limited samples are rarer \citep[e.g.][]{Schoier2001A&A...368..969S,Olofsson2002A&A...391.1053O,Ramstedt2006A&A...454L.103R}.
Such samples are essential to robustly estimate the total return of gas to the ISM from evolved stars and their lifetimes, and the large samples are useful for other scientific objectives such as studying the relationships between pulsations, dust formation and mass loss.

However, continuum studies have been more limited, with the relatively weak dust emission more difficult to observe.
Recently, \citep{Ladjal2010A&A...513A..53L} and \citep{Dharmawardena2018} have published the results of observing small samples of evolved stars using APEX/LABOCA \citep{LABOCASiringo2009A&A...497..945S} and JCMT/SCUBA-2 \citep{Holland2013} respectively.
Both studies detect bright continuum emission from the sources observed.
However, where the low mapping speed of LABOCA was insufficient to resolve the envelopes \citep{Ladjal2010A&A...513A..53L}, \citep{Dharmawardena2018} were able to detect extended emission throughout the sample with SCUBA-2, which is able to reach the same depth in roughly 1 per cent of the time.
This extended emission traces the historic mass loss, and hence the variation of mass loss on timescales of centuries, and almost all sources appear to be inconsistent with a constant mass-loss rate \citep{Dharmawardena2018}.

Motivated by the sensitivity of SCUBA-2 and the need to compare volume-limited samples of Galactic sources to the extragalactic samples studied with e.g. {\it Spitzer}, the Nearby Evolved Stars Survey (NESS) was initiated at the JCMT. 
NESS is exploiting both heterodyne and continuum observations to constrain the total gas- and dust-return rates to the Solar Neighbourhood, the dust-to-gas ratios of the outflows, the importance of cold, historic mass loss, and the physical processes driving mass loss, for a statistical sample of $\sim$ 400 nearby evolved stars.
NESS has since expanded to include the Southern sky, using APEX, and lower frequency lines, using the Nobeyama 45-metre telescope.
Preliminary results from the JCMT are favourable, with a high detection rate in both lines and continuum ($\gtrsim 75\%$, Scicluna et al., in prep; Wallstr\"om et al., in prep).

While NESS is observing a large sample at relatively low resolution to obtain a statistical picture, it is important to complement this with higher-resolution studies.
For example, Atacama Large Millimeter/sub-millimeter Array (ALMA) observations at $\sim$\,mas resolution are exploring which molecules are involved in dust formation \citep[e.g.][]{Kaminski2016A&A...592A..42K} while the new extreme adaptive optics instruments are able to trace the distribution of the newly-formed dust \citep[e.g.][]{Ohnaka2016A&A...589A..91O}.
Combining these with the large database of mass-loss rates provided by NESS will enable an exploration of the relationships between mass loss and the physics and chemistry of mass loss, by providing a robust sample for which the density, momentum and chemistry of the outflow is known.
This will help to put studies observing smaller samples in greater detail, such as the ALMA projects DEATH STAR and ATOMIUM in perspective.


The keys to observing extended emission from evolved stars are sensitivity and dynamic range; the typical evolved star has a bright, point-like (to the JCMT) central component, surrounded by a faint halo of emission that extends to very large radii \citep[e.g.][find radii up to an arcminute, and results from {\it Herschel} suggest that higher sensitivity would most likely detect emission at larger distances]{Dharmawardena2018}.
Successful recovery of this halo requires that the observation be sensitive enough to identify it, but it must also not be filtered out by observing or data-reduction techniques; chopping and high-pass filtering remove unwanted background or atmospheric emission, but also filter out interesting astronomical signals if they are on the wrong size scale.
Therefore, improved techniques for the recovery of faint emission on intermediate angular scales will be key to continuing these studies in future.
While recent projects have made advances in ensuring that all the flux is recovered from a source of this kind \citep[e.g. JINGLE, ][]{Smith2019MNRAS.486.4166S},  this remains an intensive process with many open problems.

\section{The next decade}
\label{sec:future}
In spite of ongoing progress, many questions remain open.
In this section, we explore some of these issues and how current and future instrumentation in the sub-mm can contribute to them.

\subsection{Current facilities}
\paragraph{Enrichment}

While NESS provides the data to estimate the total mass return to the Solar Neighbourhood, and hence the fuel for future star formation, to understand the chemical evolution of galaxies, it is critical to understand the origins of heavy elements.
Different mass-ranges of evolved stars will undergo significantly different nucleosynthetic processes, returning different ranges of metals to the ISM \citep[][provide a good review relevant to this section]{Karakas2014PASA...31...30K}. 
For example, the triple-$\alpha$ process produces carbon-12, and dredge-up from C-rich AGB stars is expected to be a major contributor to the carbon budget, particularly at low metallicity when C-stars are more numerous.
On the other hand, Hot Bottom Burning (HBB) converts $^{12}$C into $^{13}$C and N, reducing C production for massive AGB stars and increasing surface N.
Finally, rotational mixing in massive stars enriches the surfaces of RSGs and Wolf-Rayets (WRs) with fusion products from a variety of processes.
All of these elements are then returned to the ISM through winds.

Hence, in order to understand the enrichment of the ISM by evolved stars, we must understand not only how much each star contributes, but how the mass-loss rate changes and how the composition of the ejecta evolves with time.
The large NESS sample can begin to answer the first of these two unknowns through its large, volume-limited sample by constraining the lifetimes of the different phases, however for the shortest phases, which may be disproportionately important, more sources are required to avoid the problem of small-number statistics. 
Doubling the NESS sample, corresponding to an increase in distance of $\approx \sqrt{2}$, would provide substantially larger samples of less common objects, including C-stars, S-stars and RSGs without a significant increase in integration time, providing the data required to crack this problem.


To explore the evolution of the outflow composition, $^{13}$CO provides a useful proxy \citep[e.g.][]{GreavesHolland1997}, as the $^{13}$C abundance is sensitive to both of the key AGB nucleosythetic processes, the triple-$\alpha$ process and HBB.
However, the lower abundance compared to $^{12}$CO necessitates longer integration times to achieve the same SNR \citep[e.g. compare the $^{12}$CO and $^{13}$ CO observations of][]{DeBeck2010A&A...523A..18D}.
However, the new receivers of Namakanui provide a substantial improvement in sensitivity for compact sources compared to RxA3m and HARP, providing the means to observe large samples to significantly greater depths.
This will provide the means to achieve constraints on the isotope ratio across the entire NESS sample, revealing the evolution of the $^{13}$C abundance. 
Furthermore, new large single-dish telescopes such as the Large Millimeter Telescope (LMT), or future projects such as AtLAST, provide the further sensitivity improvements required to probe weaker isotopologues to trace the evolution of other elements that play roles in AGB nucleosynthesis, such as C$^{18}$O or HC$^{15}$N.








\subsection{Future developments}
While the size of the NESS sample provides a useful statistical overview, it is fundamentally limited in its ability to answer key questions by the nature of the observations. 
NESS is collecting sub-mm photometry for many sources, but only a subset of observations were designed to detected the extended emission associated with historic mass loss.
As intrinsically variable sources, it is also important to consider the changes in the sources at all wavelengths, requiring many observing periods. In particular, the variations are driven by pulsations, which are believed to play an important role in initiating the mass loss. 
And magnetic fields, whose importance to various classes of evolved stars has been debated over the years, remain poorly understood.
However, with current instrumentation, exploring these issues for large samples would be prohibitively expensive. 

The proposed future instrument is expected to provide the same rms noise at 850\,$\mu$m in roughly one tenth the time SCUBA-2 can achieve. 
Thanks to its intrinsic polarisation sensitivity, this increases to a factor of 20 improvement for polarimetry.
These improvements are driven by improved per-pixel sensitivity, higher detector yield, and larger field of view.
In the future, a similar improvement is projected for 450\,$\mu$m as well, either through a second instrument or an upgrade of the first.
This order-of-magnitude improvement will be transformative for several areas of evolved-star science, and we will examine the potential impact of these and other potential developments below, along with how it fits in with other facilities.

\paragraph{Mass-loss histories}
Studying the dust mass-loss histories of AGB stars is a key goal of NESS.
However, due to the increased depth required to consistently detect extended emission, only a subset ($\approx 10\%$) of NESS sources have been selected for this.
These sources are preferentially nearby ($d \leq 300$\,pc) with moderate mass-loss rates; not only is the sample barely large enough to develop a statistical picture, but it is likely to be biased as well. 
While other sources may also show resolved emission, these will tend to be the brightest sources with the highest mass-loss rates where less depth is required to detect the emission, injecting new biases into the sample.

An improvement of a factor of 10 in mapping speed would mean that even our shortest observations would begin to approach the confusion limit, highlighting the superior sensitivity of the future instrument.
At present, all NESS sources are observed using a single 30-minute Daisy, sufficient for photometry, while selected sources are observed a factor of two deeper (2 hours).
However, a 30-minute scan with the proposed camera would provide sensitivity equivalent to 5 hours with SCUBA-2, and the confusion limit would be reached in under an hour for reasonable conditions, enabling a much deeper search for extended emission than is currently possible.

\begin{figure}
    \centering
    \includegraphics[width=\textwidth]{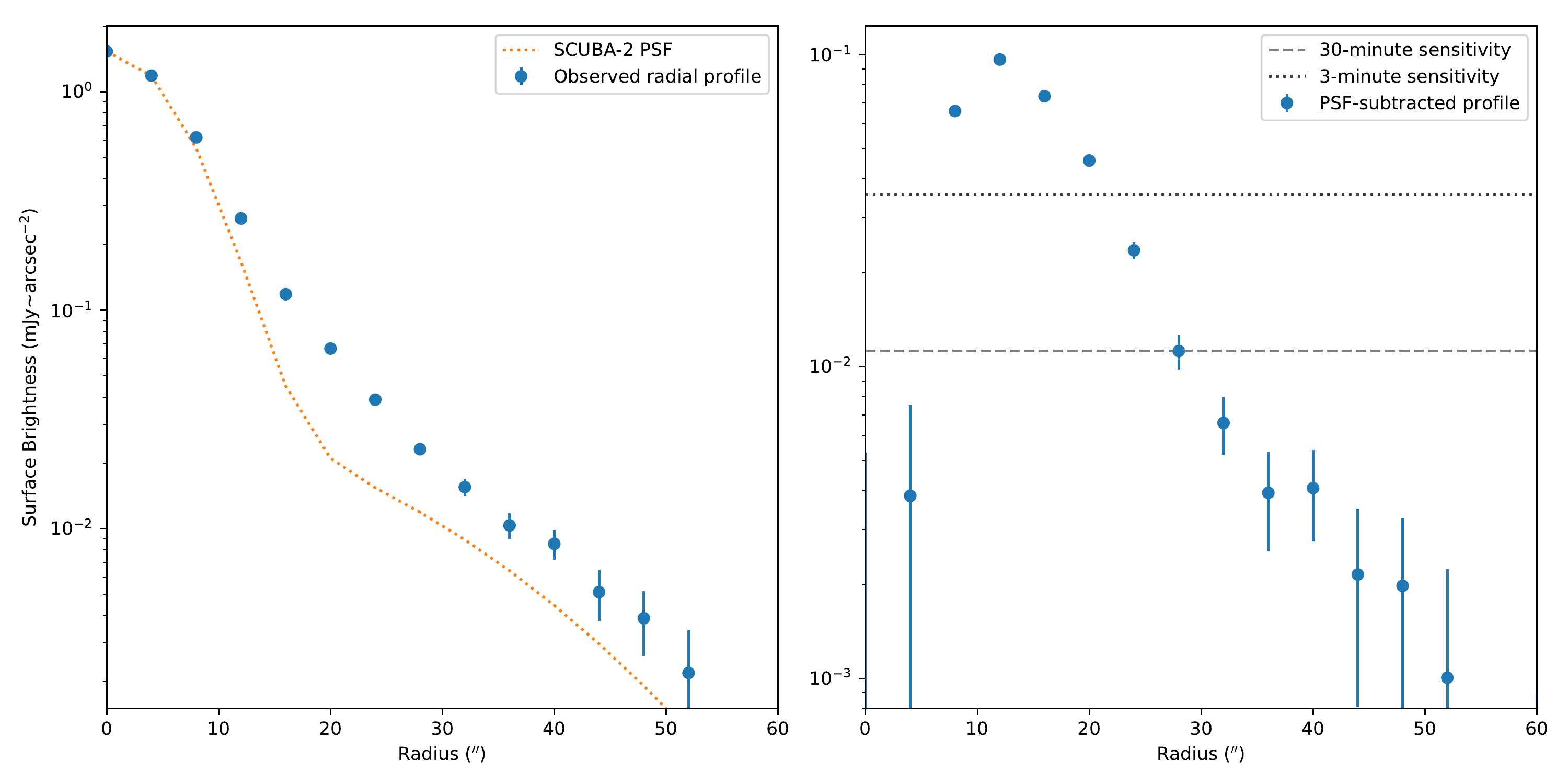}
    \caption{Deep radial profile of $o$ Cet from SCUBA-2 observations (left) and PSF-subtracted radial profile (right). This profile required 40 hours with SCUBA-2, and is expected to require $\sim$ 4 hours with the proposed 850\,$\mu$m camera. The dashed lines on the right panel indicate the expected sensitivity to extended emission for 30 and 3 minutes of integration with the new camera, and the regions of extended emission to which they would be sensitive.}
    \label{fig:oCet}
\end{figure}

An idea of what can be obtained can be seen from observations of the archetypal source $o$ Cet (Mira), which has a moderate mass-loss rate \citep[$\sim 2.5 \times 10^{-7}$\,M$_{\odot}$\,yr$^{-1}$,][]{DeBeck2010A&A...523A..18D} and is rather nearby ($\sim$ 100\,pc.
Because it is used as a pointing source, \citep{Dharmawardena2018} were able to compile all SCUBA-2 observations to provide $\approx$ 40 hours of integration.
The resulting radial profile can be seen in Fig.~\ref{fig:oCet}.
Such an observation could be repeated in just 4 hours with the proposed instrument -- given that this would allow for a better choice of observing conditions, this might even result in deeper data. 
Also indicated on the plot are the expected depths that would be achieved with the new instrument in 30 minutes or 3 minutes (the latter is equivalent to 30 minutes with SCUBA-2), under the assumption that noise scales with $\sqrt{t_{\rm int}}$ and including the advantages of azimuthal averaging.
It is immediately clear that the new camera will reveal a vastly larger area around any individual star than SCUBA-2 is able to detect, with a $\sim$50\% increase in the radius for a single scan.
This corresponds to a longer look-back time, giving a longer history of mass loss.
Along with probing a longer history for any given source, this will expand the sample of stars amenable to having their mass-loss history studied in dust continuum in two ways: by being sensitive to fainter extended emission, it will enable to study of sources with lower surface brightness (i.e. lower mass-loss rates); by probing longer timescales, sources at larger distances (harder to resolve) will become feasible.
A future large program comparable to NESS could provide mass-loss histories for all 400 stars in roughly 200 hours.
For the nearest AGB stars in such a sample ($\sim 60$\,pc), the resolution of the JCMT would correspond to a timescale of $\sim$500\,yr, while for the furthest stars ($\sim 2$\,kpc) we could expect to recover emission covering timescales up to 100\,000\,yr ($\sim 90^{\prime\prime}$).

Beyond improved sensitivity, the new camera will provide several other advantages to the study of mass-loss history. 
The improved stability of MKIDs vs. TESs is expected to improve the fidelity of images and our ability to combine multiple scans.
This would give a corresponding improvement in our confidence of the reality of structures, particularly low-surface-brightness emission, which might be further aided if increases in the field-of-view provide improvements in the estimation of the background and recovery of total power.
This will be key to interpreting mass-loss histories and, potentially, asymmetry in AGB stars, but also depends strongly on the eventual data-reduction method employed.
As noted above, SCUBA-2 data reduction becomes difficult when the required dynamic range is high or the source is faint but extended, and it will be important to mitigate these issues for evolved-star science in future.



While improved sensitivity will probe longer timescales than SCUBA-2 is able to, the only way to probe shorter timescales in the continuum is through higher resolution. 
A comparable improvement in sensitivity at 450\,$\mu$m would probe variations on timescales a factor $\sim$2 shorter, particularly in the bright inner outflow.
A byproduct of the addition of 450 would be to probe the wavelength-dependence of the emission ($\alpha$), which is primarily determined by the size and composition of dust grains.
The ability to map variations in $\alpha$ would probe the mechanisms driving the variable mass loss, by revealing whether dust properties change as a result.
While 450 offers an optimal choice of resolution and available time, other, longer wavelengths could provide comparable insights.
In particular, there are suggestions that C-stars have a significant flux excess at wavelengths $>500\mu$m, but the origin of this excess remains elusive.
Whether this represents a change in the wavelength dependence of the dust opacity or an additional population of dust requires further observations at longer wavelengths.
Observations in the 1 -- 3\,mm range, particularly in regions that avoid contamination from CO or other bright lines, are essential to determine whether this is a bump or a double power-law.
While the longer part of this range is perfectly suited to 50\,m-class facilities such as the LMT thanks to their higher resolution, the ability to observe at e.g. 1.1\,mm with the JCMT may prove particularly interesting, especially in worse weather.

An alternative way of probing shorter timescales (rather than shorter wavelength) is to combine JCMT observations with data from the SMA or ALMA.
While {\it uv-}filtering prohibits large-scale mapping with interferometers, the combination of sensitive interferometric mosaics with JCMT maps will enable the recovery of both large- and small-scale structures.
By tuning the resolution of the interferometric data, this could help to homogenise the shortest timescales probed by the maps, rather than losing the shortest timescales in more distant sources.

\begin{figure}
    \centering
    \includegraphics[width=0.8\textwidth]{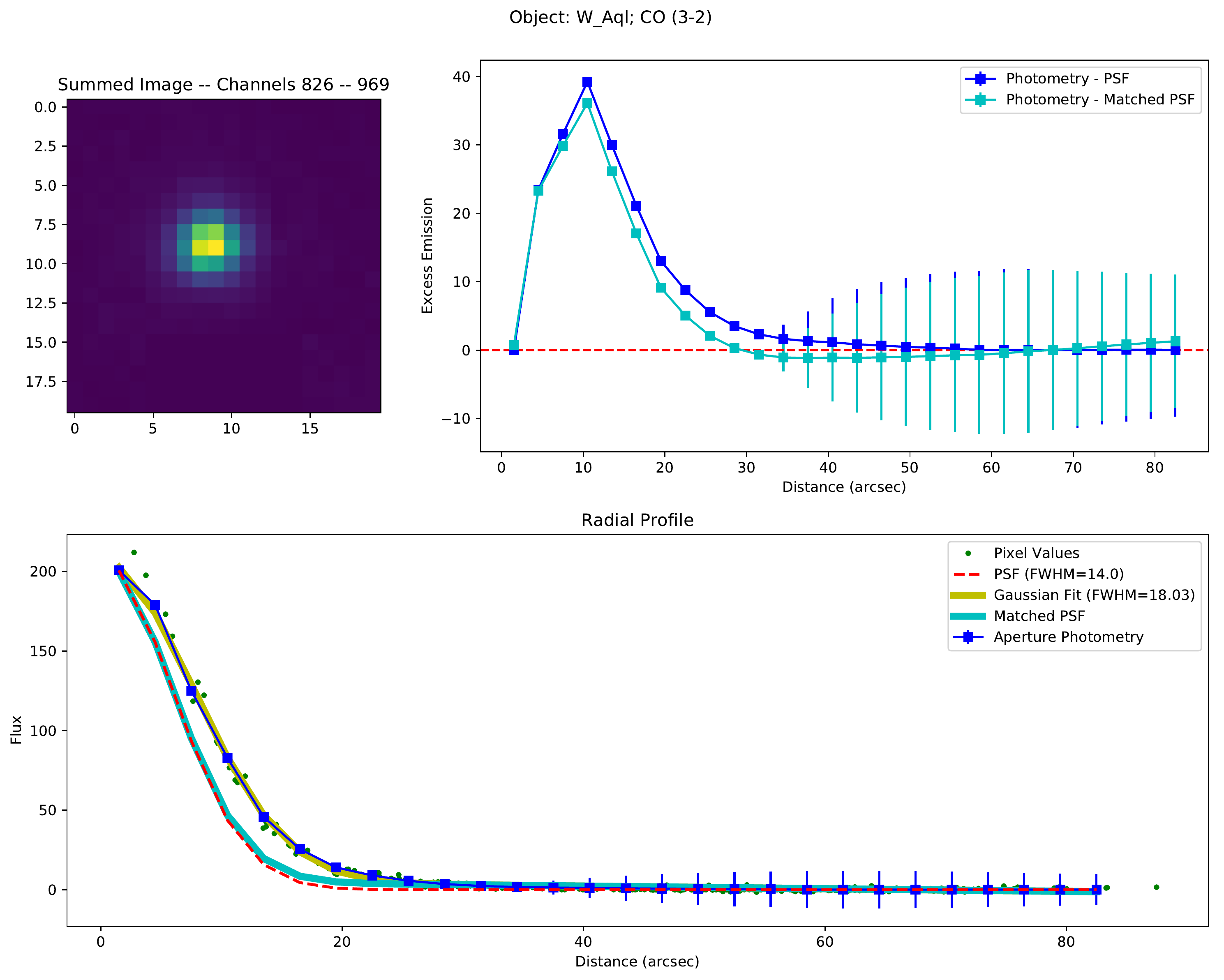}
    \caption{Example of CO map from JCMT observations, revealing significant extended emission that is likely linked to gas from historic mass loss.}
    \label{fig:COMap}
\end{figure}

While the continuum is easy to trace, dust makes up only a small fraction of the outflowing material, and it will be difficult to distinguish between variations in mass loss and variations in dust-to-gas ratio from the continuum alone.
It is therefore vital to have comparable probes of the mass-loss history from lines.
As with the continuum, the gas mass-loss history can be probed by mapping low-excitation line emission from the outer envelope (see e.g. Fig.~\ref{fig:COMap}, which shows a preliminary reduction of NESS data for this purpose showing extended CO emission out to $\sim30^{\prime\prime}$); this has certain advantages for comparison with the continuum, primarily that by tracing the cold gas the results are directly comparable to the cold dust revealed in the continuum, and that these low-J lines are sensitive to the external photodissociation of the envelope.
This has the added benefit of allowing the line contribution to the continuum flux to be determined, which has been shown to be small but have a  significant impact on the interpretation of extended continuum emission \citep[e.g.][]{Dharmawardena2019b}.

However, since different lines are primarily emitted in different regions of the outflow thanks to the differences in excitation temperature of different lines, gas mass-loss histories can also be explored through the use of multiple lines \citep[e.g.][]{Kemper2003A&A...407..609K, Decin2007A&A...475..233D}.
The bright lines of CO, HCN and their isotopologues are particularly useful for this, and the limitation of angular resolution can be circumvented by going further up the rotational ladder.
A number of useful lines lie within atmospheric windows, primarily the (4--3), (6--5) and (7--6) lines of CO, which can be observed in ALMA bands 8, 9 and 10 respectively.
Previous studies have used these lines along with the lower-J lines to infer multiple epochs of mass loss on timescales of centuries \citep{Decin2007A&A...475..233D}.
While these higher-frequency lines require good weather to observe them, the highly efficient flexible scheduling enjoyed by the JCMT makes this one of the few sites where this would be a reasonable proposition, and at present the only site in the Northern Hemisphere.
However, in future the Greenland telescope may provide an alternative site for these THz windows \citep[e.g.][]{GLTScience}.

\paragraph{Sub-mm variability}
Evolved stars are intrinsically variable, with bolometric pulsation amplitudes as high as a factor of 2.5 \citep{Wood2010}.
This variability initiates the mass loss as the pulsations launch material on ballistic trajectories suitable for dust to form; the dust is then able to provide sufficient radiation pressure to accelerate the outflow to escape \citep[e.g.][and references therein]{Liljegren2016}.
The shocks and changes in luminosity created by these pulsations sculpt the inner outflow in a number of ways: the high densities associated with the shocks drive dust formation, while the intense radiation field near maximum light destroys some of the closest dust \citep[e.g][]{Liljegren2016}; energy from shocks ionises a substantial fraction of material in the inner envelope, leading to free-free emission from the radio photosphere, while also fuelling a number of chemical pathways \citep{Menten2006,Cherchneff2006}; and the changes in luminosity lead to large changes in the pumping of high-excitation lines and hence the energy balance of the circumstellar gas \citep{He2017,He2019}.

Historically, this variability has typically been explored in the optical and near-infrared, however longer wavelengths provide a number of advantages.
The sub-mm is free from extinction, and avoids confusion caused by changes in spectral type by primarily detecting the circumstellar emission.
Furthermore, it has the potential to directly probe the influence of the variations on the outflow, rather than having to infer them indirectly: radio photospheres, dust and molecules all contribute to the sub-mm emission and its variability and their role can be traced through different observables.
By studying variability in the sub-mm and relating that to the behaviour of the stars themselves (as probed in the optical and near-IR) we can unravel the influence of the pulsations on the inner envelope, where the outflow is launched.

A preliminary study \citep{Dharmawardena2019a} exploring the two nearby AGB stars used by SCUBA-2 as pointing sources (IRC+10216 and $o$ Cet) found variability at 850\,$\mu$m in both sources of factors $\geq 1.6$ on periods that agree with those found in the optical; while the sub-mm lightcurve of $o$ Cet is in phase with the optical, IRC+10216 was out of phase, showing better agreement with the cm-wave lightcurve of \citep{Menten2006} for reasons that are not yet clear. 
However, this exploited some 40 hours of short calibration observations spread over several years for each source to construct the lightcurves.
These observations were all in the regime where the photometric uncertainty is dominated by the uncertainty in the absolute flux calibration of SCUBA-2; assuming that the calibration uncertainty for the new instrument is similar to that of SCUBA-2 ($\sim$8\%) there are at least 50 sources in the NESS sample that would be similarly calibration-dominated in 3 minutes of integration with the new instrument, and likely a substantial population of other sources similarly amenable to observation.
For such short integration, a total of 1 hour of integration is required to detect the source in 20 epochs, sufficient to evenly sample the lightcurve if a good estimate of the period is available from short wavelengths.
This suggests that samples of several tens of sources each could be feasibly explored in PI time, while a large program could potentially follow a sample of hundreds of evolved stars over a window of several years.
This will provide unprecedented insights on the behaviour of the inner envelopes of evolved stars in the sub-mm.
It will be particularly interesting to combine sub-mm monitoring with very high angular resolution snapshots of the inner envelope, for example mid-infrared interferometry or optical polarimetric imaging, which are also sensitive to the formation and properties of dust, and radio VLBI, which is dominated by the radio photosphere.
This would reveal the relative importance of these two mechanisms, probing the importance of dust formation and shock heating across the pulsation cycle.


However, such studies depend critically on the calibration uncertainty.
Given the current calibration of SCUBA-2, there is clearly a significant sample of sources that could be observed. Should the flux ratio seen in IRC+10216 and $o$ Cet prove typical, a large number of sub-mm variables could be analysed, as 8\% uncertainty would be sensitive to flux ratios between two epochs of $1.35$ -- although this ignores the statistical gains from sampling over the full lightcurve.
Further improvements to the calibration uncertainty would reveal smaller levels of variability, but require more time per epoch or a smaller sample.
Should calibration uncertainty reach just 5\%, flux ratios as low as 1.2 become observable. 



Once again, a multi-wavelength capability is important, and particularly useful if it is simultaneous, giving the ability to probe phase differences between the different wavelengths.
The spectral behaviour of the variability is able to distinguish between different sources of emission, which may reveal the underlying mechanism driving the variations.
The exact choice of band is not particularly important, although more is always better: with mechanisms such as free-free typically being more important at longer wavelengths and dust in the sub-mm, having three or more wavelengths makes it possible to distinguish between e.g. a change in the relative importance of dust and free-free or a change in the sub-mm properties of the dust.
The upcoming MKID array for the LMT, TolTEC \citep[e.g.][]{toltec}, will provide a complementary capability by mapping the 1 -- 2\,mm range at high efficiency in 3 bands; combined observations between the JCMT and LMT will break these degeracies, but the 850\,$\mu$m window is expected to be most sensitive to the dust.


Complementary to the continuum, recent studies have found significant variability in sub-mm lines, particularly those with high excitation temperatures \citep[e.g.][]{He2017,He2019}.
The variability of some of these lines correlates strongly with the changes in luminosity (probed by the near-IR), while others are strongly anti-correlated.
This may be tracing changes in thermal and radiative excitation in the inner envelope as the star pulsates.
However, many of these lines are weak, and although there have been some successful observations with APEX, such monitoring remains challenging.
Very wide bandwidth, enabling the observation of many lines simultaneously, is useful as it both maximises the integration time per spectral point and enables robust relative calibration of the lines.
Alternatively, the greater collecting area and lower beam dilution of the SMA, Morita Array or large single dishes such as the LMT may fill an important niche here.


\paragraph{Polarisation and Magnetic Fields}
Despite being fully convective and probably slow rotators, a number of studies have found significant magnetic fields in evolved stars, both at their surfaces \citep{Lebre2014A&A...561A..85L} and in their envelopes \citep[e.g.][]{Herpin2006A&A...450..667H, Vlemmings2011ApJ...728..149V}.
These have typically relied upon Zeeman splitting of either the photospheric or maser lines, which may therefore be biased as they sample only a small fraction of the emission; photospheric lines may be affected by the presence of starspots, whose lifetimes are similar to the rotation period making it difficult to sensibly average them out, while masers naturally sample dense clumps in the inner outflow.
To see the big picture and understand the typical behaviour of magnetism in evolved stars, we must observe the large-scale field in the envelope.
This requires us to map the field throughout the envelope, most easily achieved in the far-IR and sub-mm.

A commonly used tracer of the magnetic field orientation is continuum polarisation at far-IR and submm wavelengths. This arises from the thermal emission of non-spherical dust grains whose angular momenta are aligned. In the ISM, grains typically align with magnetic field lines, so that the observed polarisation angle traces the magnetic field projected on the plane of the sky 
\citep{Hildebrand1988QJRAS..29..327H, PlanckiXIX2015A&A...576A.104P, Andersson2015ARA&A..53..501A}.
However, the envelopes of AGB stars are faint in dust, requiring deep integrations to map their extended envelopes.
To date, few studies have succeeded in detecting continuum polarisation from evolved stars, and it is only with the exquisite sensitivity of ALMA that the bright RSG VY\,CMa has been detected \citep{Vlemmings2017A&A...603A..92V}.  Furthermore, as a rare case, Zeeman splitting of SiO transition is detected towards VY\,CMa \citep{Shinnaga2017PASJ...69L..10S}.
Based on the deep radial profile of IRC+10216 (the brightest AGB star in the sub-mm) from \citep{Dharmawardena2018}, POL-2 is expected to need $\sim$ 30 hours of band-1 weather to detect 5\% polarisation in the outer envelope. 
Therefore, while observations of a few sources might be feasible with SCUBA-2, a 20$\times$ increase in speed would be transformative to these studies.
The brightest sources would be feasible to observe in 1 -- 2 hours each, including 7 of the 15 sources observed by \citep{Dharmawardena2018}.
For the big picture, it would be feasible for a potential large program to observe scores or hundreds of evolved stars and map the polarisation in their envelopes, particularly if informed by the results of NESS or a future effort to map the outflows in the continuum with the proposed 850~$\mu$m instrument.

The currently favoured theory of grain alignment is that of radiative torques, or RATs, according to which the alignment of grains with the magnetic field is driven by the torque of an anisotropic radiation field on an asymmetrical (non-zero helicity) grain \citep{Dolginov1976Ap&SS..43..291D,Lazarian2007MNRAS.378..910L,Hoang2016ApJ...831..159H}. In the case of a weak magnetic field and/or a strong radiation field, and depending on the grains' properties, they may be aligned with the radiation field rather than the magnetic field \citep{Lazarian2007MNRAS.378..910L,Takazi2017ApJ...839...56T}. The presence of iron inclusions, which make the grains superparamagnetic, improves the efficiency of the alignment with the magnetic field \citep{Mathis1986ApJ...308..281M, Hoang2016ApJ...831..159H}. While RATs are currently favoured by observations \citep{Andersson2015ARA&A..53..501A}, observational constraints on this model are for the moment mostly qualitative. Studying them in the well-constrained radiation field of evolved stars provides an opportunity to better understand RATs and dust formation at the same time. 

Observations of continuum polarisation would probe a number of fundamental questions regarding evolved stars. 
How many of them have large-scale magnetic fields? What is the typical geometry?
There is some debate as to the origin of the fields, whether they are produced by angular-momentum transfer from a companion or if another mechanism is required to explain their existence.
Statistics of the presence and morphology of the large-scale field will be a key observable, as they can be compared to models of the expected population of companions, and with known binary stars.
Furthermore, comparison between the field morphology and the mass-loss history will reveal whether magnetic fields play any role in shaping the outflow.
Finally, correlating the magnetic field properties with the evolutionary stage of a sample will explore how fields evolve with the stars, and if they play a role in the changes that occur as stars evolve off the AGB.

Observations of polarisation will also have important implications for studying the mass-loss process itself. 
For evolved stars losing significant amounts of mass the outflows are expected to be driven by radiation pressure on dust that forms close to the star, fuelled by the pulsating atmosphere of the star. 
As a result, the properties of the dust that forms should have a strong influence on the properties of the outflow, such as the velocity and outflow rate.
Models of dust formation in the outflows of cool evolved stars typically assume the grains to be compact, uniform spheres, however this is unlikely to be the case. 
Deviations from symmetry have significant impacts on the absorption and scattering cross-sections \citep[e.g.][]{Siebenmorgen2014A&A...561A..82S}, and hence the radiation pressure. 
Similarly, one might also expect changes in the efficiency of gas-grain coupling. 
Fortunately, the shape -- or more precisely, the axial ratio -- of the grains is one of the main factor influencing the sub-mm polarisation fraction, making this a relatively straightforward property to explore.

In addition, there are interesting consequences for studying dust formation and processing. As mentioned above, grain alignment with magnetic fields is boosted by inclusions that makes grains superparamagnetic and, when the grains are rotating fast enough, exert a torque to align the grains with the field lines.
In AGB stars, the most likely candidate for this is iron, which in O-rich stars is expected to be a significant component of the dust from both abundance and energy-balance arguments, although the state of the iron (metallic, oxide or in silicates) remains unknown.
However, in C-rich stars, it is unclear whether iron condenses into dust as metal, ferruginous compounds or not at all.
Hence, the coupling of the dust with the magnetic field will probe whether iron is incorporated into dust grains, probed through the morphology of the polarisation.
The naive assumption is that there should be a systematic difference between O-rich and C-rich sources, with O-rich sources showing magnetic alignment, and C-rich radial alignment. 
If the polarisation pattern around around O-rich and C-rich stars were more complex than that, this could provide an opportunity to better test the RATs model itself. For instance, a shift from radial alignment to magnetic field alignment beyond a certain radius would provide better constraints on the environmental conditions needed for the two alignment regimes. Evolved stars, where the radiation field is central and has a well-understood relation with radial distance, could provide better constraints on this than molecular clouds. 


Whether or not -- and how -- iron is incorporated into dust grains directly probes the conditions in the dust-formation region, as the temperature and pressure in that zone determine the composition and size of the grains that form.
Differences in composition also have a major impact on the wavelength dependence of the polarised emission in the sub-mm \citep{Draine2009ApJ...696....1D, Guillet2018A&A...610A..16G}.
However, unlike other observables -- which are influenced not just by composition, but by size, shape, alignment efficiency and magnetic-field morphology -- the wavelength dependence of the polarisation fraction cancels out all the degeneracies: the orientation of the average magnetic field, the turbulent component of the field and, eventually, imperfect grain alignment affect polarisation in the same way independently of wavelength \citep{Greenberg1968nim..book..221G, Lee1985ApJ...290..211L}. This leaves the wavelength dependence of polarisation as a direct probe of the properties of the dust itself \citep[e.g.][]{PlanckiXXI2015A&A...576A.106P}.
Hence, multi-wavelength continuum polarimetry has an important role to play in understanding the properties of dust in evolved stars, although as yet no models exist to explore this possibility.
While TolTEC at the LMT would provide a powerful complementary capability in polarisation, our ability to map extended dust emission at the longer wavelengths that it is sensitive too has yet to be thoroughly tested.
On the other hand, \citep{Dharmawardena2018} have already demonstrated efficient mapping of the 450\,$\mu$m emission on scales only moderately smaller than at 850\,$\mu$m.
Hence, a complementary 450\,$\mu$m capability at the JCMT would provide an efficient and low-risk way to probe the wavelength dependence of polarisation.

While ALMA or the SMA would be able to image polarisation at multiple wavelengths with higher resolution, it is presently unclear whether mosaicing of polarimetric observations is feasible.
However, it is important to observe as far out in the shell as possible to minimise projection effects, which may introduce degeneracies between the 3D structure of the magnetic field and the alignment efficiency or changes in the dust properties with radius.
Hence, being able to observe the largest possible fields -- where the JCMT excels -- is important to minimising the biases in polarisation studies.


In the case of continuum polarisation, there is a key degeneracy between the magnetic field and alignment. 
For example, in the case of no polarisation, it is unclear whether there is no magnetic field, the grains are spherical or they are simply not aligned.
In the case of evolved stars, the last issue can probably be ignored -- the radiation field is so strong that RATs will spin them up to high velocities and align them with the radiation field if they are non-spherical -- leaving two options which can be selected through observations of gas polarisation.
Observations of thermally-excited lines have previously been employed, exploiting both the Goldreich-Kylafis and Zeeman effects \citep[e.g.][]{Girart2012ApJ...751L..20G, Duthu2017A&A...604A..12D}, to explore the magnetic field strength and morphology including velocity information, breaking the aforementioned degeneracy.
Hence, the ability to map polarisation with heterodyne instruments will provide an important route to constraining the large-scale magnetic fields of evolved stars, and be highly complementary to studies of the continuum polarisation.

\section{Summary}
Evolved stars have been studied in the sub-mm, and particularly with the JCMT, for many decades, primarily for their role in providing chemical and material feedback to the ISM.
Nevertheless, many important questions remain unsolved, and thanks to new instrumentation the JCMT will play a key role alongside other current and future facilities.
This white paper has presented a number of science cases where future instrumentation at the JCMT, particularly a proposed 850\,$\mu$m continuum camera which would have a factor of 10 improvement in mapping speed, can make significant contributions to the field, focusing on the study of the time-variation of mass-loss, the origins of sub-mm variability and the role of magnetic fields in mass loss.
These science goals will make key contributions to our understanding of the enrichment of the ISM with heavy elements and the physics that drives the mass-loss process.

The three key science cases we have proposed can all be achieved with a moderate time investment, thanks to the expected factor of 10 improvement in mapping speed along with necessary upgrades to the data reduction software (e.g. improving dynamic range, reducing large-scale structure filtering).
Mapping the extended emission, and hence mass-loss histories, of a sample of 400 evolved stars could be achieved in 200 hours of band 2 weather, reaching depths of roughly 10 $\mu$Jy arcsec$^{-2}$ in azimuthally-averaged profiles.
This would provide a statistical overview of the long (500 -- 100\,000 yr, depending on distance) timescale variations in mass loss, revealing their importance to mass loss and enrichment.
Monitoring the brightest 100 of these sources could, provided that the optical periods are known, probe sub-mm variability at the level of tens of per cent in 100 hours of observations in band 2 weather, split into many short observations spread over months or years, depending on the periods of individual sources.
This will reveal the mechanisms driving sub-mm variability in evolved stars and trace processes involved in dust formation and the formation of the radio photosphere.
A sample of 100 bright, extended evolved stars could be mapped in polarisation in 200 hours of band 1 weather to a depth required to detect 5\% polarisation in their outer envelopes.
By providing a statistical overview of the morphology of polarised emission in evolved stars, this would answer key open questions about the role of magnetic fields in shaping their outflows, the composition and shape of dust (with important consequences for driving the outflow with radiation pressure) and the physics of grain alignment.

Mapping in the sub-mm, where dust dominates the continuum in evolved stars, is a unique niche for the JCMT.
The combination of large field of view and high sensitivity of the proposed receiver provide a natural means for exploring extended emission, which is key to several future studies of evolved stars.
Observations in the sub-mm, particularly at 850\,$\mu$m are decidedly important as this probes dust temperatures in the same range as the gas temperatures probed by the low-J rotational transitions of CO, the favoured way of measuring the gas in the outflow.
Other facilities, such as the LMT, can complement the proposed studies through similar capabilities at longer wavelengths, which can help to decouple the influence of different emission mechanisms and constrain the properties of the dust.
Despite its exquisite sensitivity, the small field of view and {\it uv}-filtering mean that ALMA is unable to compete, although it can supplement the JCMT by probing variations on smaller angular scales, corresponding to shorter times.
The combination of these facilities, supporting the proposed continuum instrument, represent a bright future for studies of evolved stars in the sub-mm continuum.








\bibliographystyle{unsrtMaxAuth} 
\bibliography{references.bib}  

\begin{thebibliography}{10}

\bibitem{Karakas2014IAUS..298..142K}
Amanda~I. {Karakas}.
\newblock {Stellar yields for chemical evolution modelling}.
\newblock In Sofia {Feltzing}, Gang {Zhao}, Nicholas~A. {Walton}, and Patricia
  {Whitelock}, editors, {\em Setting the scene for Gaia and LAMOST}, volume 298
  of {\em IAU Symposium}, pages 142--153, Jan 2014.

\bibitem{Karakas2014PASA...31...30K}
Amanda~I. {Karakas} and John~C. {Lattanzio}.
\newblock {The Dawes Review 2: Nucleosynthesis and Stellar Yields of Low- and
  Intermediate-Mass Single Stars}.
\newblock {\em \pasa}, 31:e030, Jul 2014.

\bibitem{Meynet2017IAUS..329....3M}
Georges {Meynet}, Andr{\'e} {Maeder}, Cyril {Georgy}, and
  \bibinfo{person}{others}.
\newblock {Massive stars, successes and challenges}.
\newblock In J.~J. {Eldridge}, J.~C. {Bray}, L.~A.~S. {McClelland}, and
  L.~{Xiao}, editors, {\em The Lives and Death-Throes of Massive Stars}, volume
  329 of {\em IAU Symposium}, pages 3--14, Nov 2017.

\bibitem{Hofner2018A&ARv..26....1H}
Susanne {H{\"o}fner} and Hans {Olofsson}.
\newblock {Mass loss of stars on the asymptotic giant branch. Mechanisms,
  models and measurements}.
\newblock {\em \aapr}, 26(1):1, Jan 2018.

\bibitem{Kim2015}
Hyosun {Kim}, Sheng-Yuan {Liu}, Naomi {Hirano}, and \bibinfo{person}{others}.
\newblock {High-resolution CO Observation of the Carbon Star CIT 6 Revealing
  the Spiral Structure and a Nascent Bipolar Outflow}.
\newblock {\em \apj}, 814(1):61, Nov 2015.

\bibitem{Kim2017}
Hyosun {Kim}, Alfonso {Trejo}, Sheng-Yuan {Liu}, and \bibinfo{person}{others}.
\newblock {The large-scale nebular pattern of a superwind binary in an
  eccentric orbit}.
\newblock {\em Nature Astronomy}, 1:0060, Mar 2017.

\bibitem{Smith2015}
Nathan {Smith} and Ryan {Tombleson}.
\newblock {Luminous blue variables are antisocial: their isolation implies that
  they are kicked mass gainers in binary evolution}.
\newblock {\em \mnras}, 447(1):598--617, Feb 2015.

\bibitem{GreavesHolland1997}
J.~S. {Greaves} and W.~S. {Holland}.
\newblock {High mass-loss carbon stars and the evolution of the local
  $^{12}$C/$^{13}$C ratio.}
\newblock {\em \aap}, 327:342--348, Nov 1997.

\bibitem{Young1995ApJ...445..872Y}
K.~{Young}.
\newblock {A CO(3--2) Survey of Nearby Mira Variables}.
\newblock {\em \apj}, 445:872, Jun 1995.

\bibitem{Knapp1985ApJ...292..640K}
G.~R. {Knapp} and M.~{Morris}.
\newblock {Mass Loss from Evolved Stars. III. Mass Loss Rates for 50 Stars from
  CO J = 1--0 Observations}.
\newblock {\em \apj}, 292:640, May 1985.

\bibitem{Kahane1994A&A...290..183K}
C.~{Kahane} and M.~{Jura}.
\newblock {Circumstellar CO around bright oxygen-rich semi-regulars.}
\newblock {\em \aap}, 290:183--197, Oct 1994.

\bibitem{Schoier2001A&A...368..969S}
F.~L. {Sch{\"o}ier} and H.~{Olofsson}.
\newblock {Models of circumstellar molecular radio line emission. Mass loss
  rates for a sample of bright carbon stars}.
\newblock {\em \aap}, 368:969--993, Mar 2001.

\bibitem{Olofsson2002A&A...391.1053O}
H.~{Olofsson}, D.~{Gonz{\'a}lez Delgado}, F.~{Kerschbaum}, and F.~L.
  {Sch{\"o}ier}.
\newblock {Mass loss rates of a sample of irregular and semiregular M-type
  AGB-variables}.
\newblock {\em \aap}, 391:1053--1067, Sep 2002.

\bibitem{Ramstedt2006A&A...454L.103R}
S.~{Ramstedt}, F.~L. {Sch{\"o}ier}, H.~{Olofsson}, and A.~A. {Lundgren}.
\newblock {Mass-loss properties of S-stars on the AGB}.
\newblock {\em \aap}, 454(2):L103--L106, Aug 2006.

\bibitem{Ladjal2010A&A...513A..53L}
D.~{Ladjal}, K.~{Justtanont}, M.~A.~T. {Groenewegen}, and
  \bibinfo{person}{others}.
\newblock {870 {\ensuremath{\mu}}m observations of evolved stars with LABOCA}.
\newblock {\em \aap}, 513:A53, Apr 2010.

\bibitem{Dharmawardena2018}
Thavisha~E. {Dharmawardena}, Francisca {Kemper}, Peter {Scicluna}, and
  \bibinfo{person}{others}.
\newblock {Extended Dust Emission from Nearby Evolved Stars}.
\newblock {\em \mnras}, 479(1):536--552, Sep 2018.

\bibitem{LABOCASiringo2009A&A...497..945S}
G.~{Siringo}, E.~{Kreysa}, A.~{Kov{\'a}cs}, and \bibinfo{person}{others}.
\newblock {The Large APEX BOlometer CAmera LABOCA}.
\newblock {\em \aap}, 497(3):945--962, Apr 2009.

\bibitem{Holland2013}
W.~S. {Holland}, D.~{Bintley}, E.~L. {Chapin}, and \bibinfo{person}{others}.
\newblock {SCUBA-2: the 10 000 pixel bolometer camera on the James Clerk
  Maxwell Telescope}.
\newblock {\em MNRAS}, 430(4):2513--2533, Apr 2013.

\bibitem{Kaminski2016A&A...592A..42K}
T.~{Kami{\'n}ski}, K.~T. {Wong}, M.~R. {Schmidt}, and \bibinfo{person}{others}.
\newblock {An observational study of dust nucleation in Mira (o Ceti). I.
  Variable features of AlO and other Al-bearing species}.
\newblock {\em \aap}, 592:A42, Jul 2016.

\bibitem{Ohnaka2016A&A...589A..91O}
K.~{Ohnaka}, G.~{Weigelt}, and K.~H. {Hofmann}.
\newblock {Clumpy dust clouds and extended atmosphere of the AGB star W Hydrae
  revealed with VLT/SPHERE-ZIMPOL and VLTI/AMBER}.
\newblock {\em \aap}, 589:A91, May 2016.

\bibitem{Smith2019MNRAS.486.4166S}
Matthew W.~L. {Smith}, Christopher J.~R. {Clark}, Ilse {De Looze}, and
  \bibinfo{person}{others}.
\newblock {JINGLE, a JCMT legacy survey of dust and gas for galaxy evolution
  studies: II. SCUBA-2 850 {\ensuremath{\mu}}m data reduction and dust flux
  density catalogues}.
\newblock {\em \mnras}, 486(3):4166--4185, Jul 2019.

\bibitem{DeBeck2010A&A...523A..18D}
E.~{De Beck}, L.~{Decin}, A.~{de Koter}, and \bibinfo{person}{others}.
\newblock {Probing the mass-loss history of AGB and red supergiant stars from
  CO rotational line profiles. II. CO line survey of evolved stars: derivation
  of mass-loss rate formulae}.
\newblock {\em \aap}, 523:A18, Nov 2010.

\bibitem{Dharmawardena2019b}
Thavisha~E. {Dharmawardena}, Francisca {Kemper}, Sundar {Srinivasan}, and
  \bibinfo{person}{others}.
\newblock {The nearby evolved stars survey - I. JCMT/SCUBA-2 submillimetre
  detection of the detached shell of U Antliae}.
\newblock {\em \mnras}, 489(3):3218--3231, Nov 2019.

\bibitem{Kemper2003A&A...407..609K}
F.~{Kemper}, R.~{Stark}, K.~{Justtanont}, and \bibinfo{person}{others}.
\newblock {Mass loss and rotational CO emission from Asymptotic Giant Branch
  stars}.
\newblock {\em \aap}, 407:609--629, Aug 2003.

\bibitem{Decin2007A&A...475..233D}
L.~{Decin}, S.~{Hony}, A.~{de Koter}, and \bibinfo{person}{others}.
\newblock {The variable mass loss of the AGB star WX Piscium as traced by the
  CO J = 1-0 through 7-6 lines and the dust emission}.
\newblock {\em \aap}, 475(1):233--242, Nov 2007.

\bibitem{GLTScience}
Hiroyuki {Hirashita}, Patrick~M. {Koch}, Satoki {Matsushita}, and
  \bibinfo{person}{others}.
\newblock {First-generation science cases for ground-based terahertz
  telescopes}.
\newblock {\em \pasj}, 68(1):R1, Feb 2016.

\bibitem{Wood2010}
P.~R. {Wood}.
\newblock {Evolutionary and pulsation properties of AGB stars .}
\newblock {\em \memsai}, 81:883, Jan 2010.

\bibitem{Liljegren2016}
S.~{Liljegren}, S.~{H{\"o}fner}, W.~{Nowotny}, and K.~{Eriksson}.
\newblock {Dust-driven winds of AGB stars: The critical interplay of
  atmospheric shocks and luminosity variations}.
\newblock {\em \aap}, 589:A130, May 2016.

\bibitem{Menten2006}
K.~M. {Menten}, M.~J. {Reid}, E.~{Kr{\"u}gel}, M.~J. {Claussen}, and
  R.~{Sahai}.
\newblock {Radio continuum monitoring of the extreme carbon star IRC+10216}.
\newblock {\em \aap}, 453(1):301--307, Jul 2006.

\bibitem{Cherchneff2006}
I.~{Cherchneff}.
\newblock {A chemical study of the inner winds of asymptotic giant branch
  stars}.
\newblock {\em \aap}, 456(3):1001--1012, Sep 2006.

\bibitem{He2017}
J.~H. {He}, {Dinh-V-Trung}, and T.~I. {Hasegawa}.
\newblock {Monitor Variability of Millimeter Lines in IRC+10216}.
\newblock {\em \apj}, 845(1):38, Aug 2017.

\bibitem{He2019}
J.~H. {He}, T.~{Kami{\'n}ski}, R.~E. {Mennickent}, and
  \bibinfo{person}{others}.
\newblock {ALMA Monitoring of Millimeter Line Variation in IRC +10216. I.
  Overview of Millimeter Variability}.
\newblock {\em \apj}, 883(2):165, Oct 2019.

\bibitem{Dharmawardena2019a}
Thavisha~E. {Dharmawardena}, Francisca {Kemper}, Jan G.~A. {Wouterloot}, and
  \bibinfo{person}{others}.
\newblock {The sub-mm variability of IRC+10216 and o Ceti}.
\newblock {\em \mnras}, 489(3):3492--3505, Nov 2019.

\bibitem{toltec}
J.~E. {Austermann}, J.~A. {Beall}, S.~A. {Bryan}, and \bibinfo{person}{others}.
\newblock {Millimeter-Wave Polarimeters Using Kinetic Inductance Detectors for
  TolTEC and Beyond}.
\newblock {\em Journal of Low Temperature Physics}, 193(3-4):120--127, Nov
  2018.

\bibitem{Lebre2014A&A...561A..85L}
A.~{L{\`e}bre}, M.~{Auri{\`e}re}, N.~{Fabas}, and \bibinfo{person}{others}.
\newblock {Search for surface magnetic fields in Mira stars. First detection in
  {\ensuremath{\chi}} Cygni}.
\newblock {\em \aap}, 561:A85, Jan 2014.

\bibitem{Herpin2006A&A...450..667H}
F.~{Herpin}, A.~{Baudry}, C.~{Thum}, D.~{Morris}, and H.~{Wiesemeyer}.
\newblock {Full polarization study of SiO masers at 86 GHz}.
\newblock {\em \aap}, 450(2):667--680, May 2006.

\bibitem{Vlemmings2011ApJ...728..149V}
W.~H.~T. {Vlemmings}, E.~M.~L. {Humphreys}, and R.~{Franco-Hern{\'a}ndez}.
\newblock {Magnetic Fields in Evolved Stars: Imaging the Polarized Emission of
  High-frequency SiO Masers}.
\newblock {\em \apj}, 728(2):149, Feb 2011.

\bibitem{Hildebrand1988QJRAS..29..327H}
Roger~H. {Hildebrand}.
\newblock {Magnetic fields and stardust}.
\newblock {\em \qjras}, 29:327--351, Sep 1988.

\bibitem{PlanckiXIX2015A&A...576A.104P}
{Planck Collaboration}, P.~A.~R. {Ade}, N.~{Aghanim}, and
  \bibinfo{person}{others}.
\newblock {Planck intermediate results. XIX. An overview of the polarized
  thermal emission from Galactic dust}.
\newblock {\em \aap}, 576:A104, Apr 2015.

\bibitem{Andersson2015ARA&A..53..501A}
B.~G. {Andersson}, A.~{Lazarian}, and John~E. {Vaillancourt}.
\newblock {Interstellar Dust Grain Alignment}.
\newblock {\em \araa}, 53:501--539, Aug 2015.

\bibitem{Vlemmings2017A&A...603A..92V}
W.~H.~T. {Vlemmings}, T.~{Khouri}, I.~{Mart{\'\i}-Vidal}, and
  \bibinfo{person}{others}.
\newblock {Magnetically aligned dust and SiO maser polarisation in the envelope
  of the red supergiant VY Canis Majoris}.
\newblock {\em \aap}, 603:A92, Jul 2017.

\bibitem{Shinnaga2017PASJ...69L..10S}
H.~{Shinnaga}, M.~J. {Claussen}, S.~{Yamamoto}, and M.~{Shimojo}.
\newblock {Strong magnetic field generated by the extreme oxygen-rich red
  supergiant VY Canis Majoris}.
\newblock {\em \pasj}, 69:L10, December 2017.

\bibitem{Dolginov1976Ap&SS..43..291D}
A.~Z. {Dolginov} and I.~G. {Mitrofanov}.
\newblock {Orientation of Cosmic Dust Grains}.
\newblock {\em \apss}, 43(2):291--317, Sep 1976.

\bibitem{Lazarian2007MNRAS.378..910L}
A.~{Lazarian} and Thiem {Hoang}.
\newblock {Radiative torques: analytical model and basic properties}.
\newblock {\em \mnras}, 378(3):910--946, Jul 2007.

\bibitem{Hoang2016ApJ...831..159H}
Thiem {Hoang} and A.~{Lazarian}.
\newblock {A Unified Model of Grain Alignment: Radiative Alignment of
  Interstellar Grains with Magnetic Inclusions}.
\newblock {\em \apj}, 831(2):159, Nov 2016.

\bibitem{Takazi2017ApJ...839...56T}
Ryo {Tazaki}, Alexandre {Lazarian}, and Hideko {Nomura}.
\newblock {Radiative Grain Alignment In Protoplanetary Disks: Implications for
  Polarimetric Observations}.
\newblock {\em \apj}, 839(1):56, Apr 2017.

\bibitem{Mathis1986ApJ...308..281M}
J.~S. {Mathis}.
\newblock {The Alignment of Interstellar Grains}.
\newblock {\em \apj}, 308:281, Sep 1986.

\bibitem{Siebenmorgen2014A&A...561A..82S}
R.~{Siebenmorgen}, N.~V. {Voshchinnikov}, and S.~{Bagnulo}.
\newblock {Dust in the diffuse interstellar medium. Extinction, emission,
  linear and circular polarisation}.
\newblock {\em \aap}, 561:A82, Jan 2014.

\bibitem{Draine2009ApJ...696....1D}
Bruce~T. {Draine} and Aur{\'e}lien~A. {Fraisse}.
\newblock {Polarized Far-Infrared and Submillimeter Emission from Interstellar
  Dust}.
\newblock {\em \apj}, 696(1):1--11, May 2009.

\bibitem{Guillet2018A&A...610A..16G}
V.~{Guillet}, L.~{Fanciullo}, L.~{Verstraete}, and \bibinfo{person}{others}.
\newblock {Dust models compatible with Planck intensity and polarization data
  in translucent lines of sight}.
\newblock {\em \aap}, 610:A16, Feb 2018.

\bibitem{Greenberg1968nim..book..221G}
J.~Mayo {Greenberg}.
\newblock {\em {Interstellar Grains}}, page 221.
\newblock 1968.

\bibitem{Lee1985ApJ...290..211L}
H.~M. {Lee} and B.~T. {Draine}.
\newblock {Infrared extinction and polarization due to partially aligned
  spheroidal grains : models for the dust toward the BN object.}
\newblock {\em \apj}, 290:211--228, Mar 1985.

\bibitem{PlanckiXXI2015A&A...576A.106P}
{Planck Collaboration}, P.~A.~R. {Ade}, N.~{Aghanim}, and
  \bibinfo{person}{others}.
\newblock {Planck intermediate results. XXI. Comparison of polarized thermal
  emission from Galactic dust at 353 GHz with interstellar polarization in the
  visible}.
\newblock {\em \aap}, 576:A106, Apr 2015.

\bibitem{Girart2012ApJ...751L..20G}
J.~M. {Girart}, N.~{Patel}, W.~H.~T. {Vlemmings}, and Ramprasad {Rao}.
\newblock {Mapping the Linearly Polarized Spectral Line Emission around the
  Evolved Star IRC+10216}.
\newblock {\em \apjl}, 751(1):L20, May 2012.

\bibitem{Duthu2017A&A...604A..12D}
A.~{Duthu}, F.~{Herpin}, H.~{Wiesemeyer}, and \bibinfo{person}{others}.
\newblock {Magnetic field in IRC+10216 and other C-rich evolved stars}.
\newblock {\em \aap}, 604:A12, Jul 2017.

\end{thebibliography}





\end{document}